# A novel normalized sign algorithm for system identification under impulsive noise interference

Lu Lu • Haiquan Zhao • Kan Li • Badong Chen

**Abstract.** To overcome the performance degradation of adaptive filtering algorithms in the presence of impulsive noise, a novel normalized sign algorithm (NSA) based on a convex combination strategy, called NSA-NSA, is proposed in this paper. The proposed algorithm is capable of solving the conflicting requirement of fast convergence rate and low steady-state error for an individual NSA filter. To further improve the robustness to impulsive noises, a mixing parameter updating formula based on a sign cost function is derived. Moreover, a tracking weight transfer scheme of coefficients from a fast NSA filter to a slow NSA filter is proposed to speed up the convergence rate. The convergence behavior and performance of the new algorithm are verified by theoretical analysis and simulation studies.

## 1 Introduction

In general, the performance of an adaptive filtering algorithm degrades when signals are contaminated by impulsive or heavy-tailed noise. To overcome this limitation, many algorithms were proposed, such as the sign algorithm (SA) [26], the signed regressor algorithm (SRA) [7] and the sign-sign algorithm (SSA) [9]. Although the SA has been successfully applied to system identification under impulsive noise, its convergence rate is slow [26]. As a variant of SA, the convergence behavior of the SRA is heavily dependent on the inputs, and there may exist some inputs for which the SRA is unstable while the least mean square (LMS) algorithm is stable [7]. Among the family of SA algorithms, the SSA has the lowest computational complexi-

L. Lu • H. Zhao(✉)
Key Laboratory of Magnetic Suspension Technology and Maglev Vehicle, Ministry of Education, and
School of Electrical Engineering, Southwest Jiaotong University, Chengdu, China
e-mail: lulu@my.swjtu.edu.cn, hqzhao@home.swjtu.edu.cn

Kan Li
Computational Neuro-Engineering Laboratory, University of Florida, Gainesville, FL 32611, USA
email: likan@ufl.edu.

B. Chen
School of Electronic and Information Engineering, Xi'an Jiaotong University, Xi'an, China.
e-mail: chenbd@mail.xjtu.edu.cn

ty and the most similar characteristic to SA [9]. In addition, the degradations of two algorithms depend significantly on the initial weights. Similar to the normalized least mean square (NLMS), the normalized versions of these sign algorithms can be easily derived, including the normalized SA (NSA) [10], the normalized SRA (NSRA) [11] and the normalized SSA (NSSA) [12]. The NSA can improve the robustness of the filter against impulsive noises. However, its convergence performance is still not good in general. Several variants have been proposed aiming at improving the convergence [5-6,8,13-14,16,27,30,32,]. Particularly, in [14], a dual SA (DSA) with a variable step-size (VSS) was proposed, but it has a local divergence problem especially when a large disparity occurs between two successive step sizes. In [8], attempt was made to obtain better stability and convergence performance by inserting another step-size. Note that the above-mentioned efforts have all been made for a single adaptive filtering architecture.

On the other hand, to cope with impulsive noise, the family of mixed-norm algorithms were developed to combine the benefits of stochastic gradient adaptive filter algorithms [3-4,17,24,31]. Chambers *et al.* introduced a robust mixed-norm (RMN) algorithm, where the cost function is a combination of the error norms that underlie the LMS and SA [4]. Later, Papoulis et al. [17,23] proposed a novel VSS RMN (NRMN) algorithm, which circumvents the drawback of slow convergence for RMN to some extent, by using time-varying learning rate..

The convex combination approach is another way to effectively balance the convergence rate and steady-state error. An adaptive approach using combination LMS (CLMS) was proposed in [1], utilizing two LMS filters with different step sizes to obtain fast convergence and small misadjustment. Nevertheless, when the signals are corrupted by impulsive noise, the algorithms in [1] and [15] usually fail to converge. To improve performance, an NLMS-NSA algorithm was developed where a combination scheme was used to switch between the NLMS and NSA algorithms [2]. Regrettably, in the initial stage of adaptation, the NLMS algorithm may cause large misadjustment especially when the noise becomes severe. Moreover, the adaptation rule of the mixing parameter of NLMS-NSA is unsuitable for impulsive noise, such that the algorithm fails to perform at a desirable level.

In this work, to address the above-mentioned problems, a NSA-NSA algorithm is proposed by using the convex combination approach. This novel algorithm achieves robust performance in impulsive noise environments by leveraging two independent NSA filters with a large and a small step-sizes, respectively. To further enhance the robustness against impulsive noise, the mixing parameter is adjusted using a sign cost function. In addition, a tracking weight transfer of coefficients is proposed in order to obtain fast convergence speed during a transition period. Our main contributions are listed as follows: 1) propose a NSA-NSA that is well-suited for system identification problems under impulsive noise; 2) modify an existing update scheme of the mixing parameter, and analyze its behavior; 3) propose a novel weight transfer scheme that is computationally simple yet can significantly improve the convergence rate.

The rest of this paper is organized as follows. In Section 2, we propose the NSA-NSA and develop a novel weight transfer scheme. In Section 3, simulation results in different impulsive noise environments are presented. Lastly, Section 4 concludes the paper.

## 2 Adaptive combination of NSA algorithms

### 2.1 The proposed algorithm

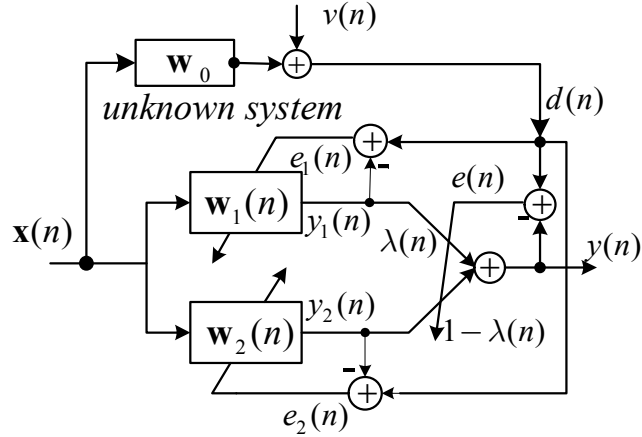

**Fig. 1.** Diagram of the proposed algorithm.

The diagram of adaptive combination scheme of two NSA filters is illustrated in Fig. 1, where $\mathbf{x}(n)$ and $y(n)$ are the filter input and output signals respectively, $d(n)$ is the desired signal, $y_1(n)$ and $y_2(n)$ are symbols of the two component filters defined by weight vectors $\mathbf{w}_1$ and $\mathbf{w}_2$, respectively, $v(n)$ is the impulsive noise, and $\mathbf{w}_0$ is the weight vector of the unknown system. The overall error of the combined filter is given by $e(n) = d(n) - y(n)$. To improve performance, both filters are combined with a scalar mixing parameter $\lambda(n)$:

$$y(n) = \lambda(n)y_1(n) + [1-\lambda(n)]y_2(n) \tag{1}$$

$$e(n) = \lambda(n)e_1(n) + [1-\lambda(n)]e_2(n) \tag{2}$$

where $\lambda(n) \in [0,1]$ is defined by a sigmoidal activation function with auxiliary parameter $a(n)$

$$\lambda(n) = 1/(1+e^{-a(n)}). \tag{3}$$

A gradient descent adaptation of $a(n)$ is given as

$$\begin{aligned} a(n+1) &= a(n) - \frac{v_a}{2}\frac{\partial e^2(n)}{\partial a(n)} \\ &= a(n) + v_a e(n)[y_1(n) - y_2(n)]\lambda(n)[1-\lambda(n)]. \end{aligned} \tag{4}$$

Note that $v_a$ is the step-size of the auxiliary parameter $a(n)$. This adaptation rule is derived by the cost function $J(n) = e(n)^2$ [1]. To improve the robustness against impulsive noise, the new cost function is defined as $J_s(n) = |e(n)|$ based on the classical sign-error LMS algorithm [26]. Therefore, the updated scheme of $a(n)$ is derived by minimizing the cost function $J_s(n)$ as follows:

$$a(n+1) = a(n) - \frac{\mu_a}{2} \frac{\partial J_s(n)}{\partial a(n)} \tag{5}$$

where $\mu_a$ is the step-size.

Using the chain rule, the gradient adaptation of $J_s(n)$ can be calculated as follows:

$$\begin{aligned} a(n+1) &= a(n) - \frac{\mu_a}{2} \frac{\partial J_s(n)}{\partial \lambda(n)} \frac{\partial \lambda(n)}{\partial a(n)} \\ &= a(n) + \rho_a sign\{e(n)\}[y_1(n) - y_2(n)]\lambda(n)[1-\lambda(n)] \end{aligned} \tag{6}$$

where $\rho_a$ is a positive constant, and the sign function $sign(\cdot)$ can be expressed as

$$sign(x) = \frac{x}{\|x\|_2} = \begin{cases} 1, & if\ x > 0 \\ 0, & if\ x = 0 \\ -1, & if\ x < 0 \end{cases} . \tag{7}$$

At each iteration cycle, the weight update of NSA-NSA takes the form [10]

$$\mathbf{w}_i(n+1) = \mathbf{w}_i(n) + \mu_i \frac{\mathbf{x}(n)sign\{e_i(n)\}}{\varepsilon_i + \|\mathbf{x}(n)\|_2^2} \quad (i=1,2) \tag{8}$$

where $\mathbf{w}_i(n)$ is the weight vectors with length $M$, $\mu_i$ is the constant step-size, $\varepsilon_i > 0$ is a regularization constant close to zero, and $\|\cdot\|_2$ represents the Euclidian-norm. As a result, the combined filter is obtained by using the following convex combination scheme

$$\mathbf{w}(n) = \lambda(n)\mathbf{w}_1(n) + [1-\lambda(n)]\mathbf{w}_2(n) . \tag{9}$$

### 2.2 Proposed weight transfer scheme

Inspired by the instantaneous transfer scheme from [22], a tracking weight transfer scheme is proposed, as shown in Table 1. By using a sliding window approach, the proposed scheme involves few parameters and retains the robustness against impulsive noise with low-cost. Like the instantaneous transfer scheme in [22], the parameter of proposed weight transfer scheme is not sensitive to the choice. This scheme can speed up the convergence property of the overall filter, especially during the period of convergence transition. Define $N_0$ as the window length. If $n-1$ mod $N_0$ is equal to zero, then implement the following operations. It is well known that the standard convex combination scheme needs to check if $a(n+1) = a^+$, so the only additional operation is the $n$ mod $N_0$ operation. The judgment condition $a(n+1) \geq a^+$ represents the condition when the fast filter (filter with large step-size) switches to the slow filter

(filter with small step size) at the transient stage. The operations $\lambda(n+1)=0$ and $\lambda(n+1)=1$ are the limitations for $a(n+1)<-a^+$ and $a(n+1)\geq a^+$, respectively. The operation $\mathbf{w}_2(n+1)=\mathbf{w}_1(n+1)$ denotes the transfer of coefficients, which is only applied in the transient stage. By applying the weight transfer, the adaptation of $\mathbf{w}_2(n+1)$ is similar to that of the fast filter, which speeds up the convergence rate of $\mu_2$ NSA filter. Moreover, the cost of the proposed weight transfer scheme is smaller than that of the original combination, because only one filter is adapted.

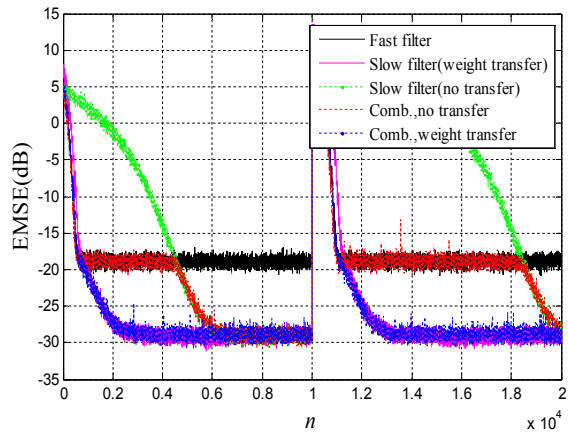

**Fig. 2.** Comparison of EMSE of NSA-NSA for Gaussian input in example 1.

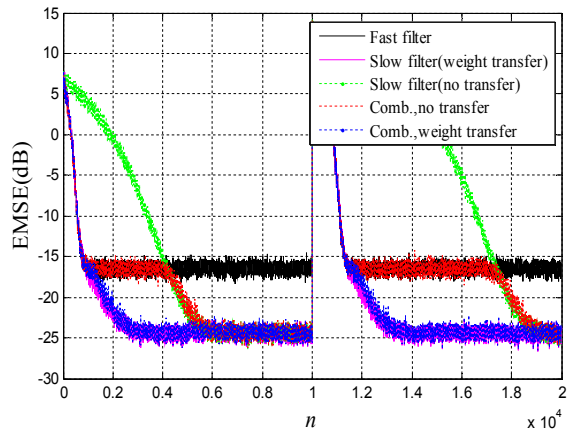

**Fig. 3.** Comparison of EMSE of NSA-NSA for Gaussian input in example 2.

Figs. 2 and 3 display the comparison of excess means-square error (EMSE) obtained from NSA-NSA with the tracking weight transfer scheme and no transfer scheme (the mixing parameter is adjusted according to (6)). The same step-size is chosen for this comparison. As can be seen, the overall performance of the filter bank is improved by the transfer scheme. It shows from these figures that the proposed weight transfer scheme exhibit faster convergence than no transfer scheme. The proposed algorithm is summarized in Table 1.

**Table 1.** The Proposed algorithm

---
**Initialize** $N_0, a^+, \rho_a, \varepsilon_i, \mu_i, \mathbf{w}_i(0) = 0, a(0) = 0, \lambda(0) = 0.5$
**Loop** $n=1 \rightarrow$ **end do**
$\quad y_i(n) = \mathbf{w}_i^T(n)\mathbf{x}(n) \quad (i = 1, 2)$
$\quad e_i(n) = d(n) - y_i(n)$
$\quad y(n) = \lambda(n)y_1(n) + [1 - \lambda(n)]y_2(n)$
$\quad \mathbf{w}_1(n+1) = \mathbf{w}_1(n) + \mu_1 \dfrac{sign\{e_1(n)\}\mathbf{x}(n)}{\varepsilon_1 + \|\mathbf{x}(n)\|_2^2}$
$\quad \mathbf{w}_2(n+1) = \mathbf{w}_2(n) + \mu_2 \dfrac{sign\{e_2(n)\}\mathbf{x}(n)}{\varepsilon_2 + \|\mathbf{x}(n)\|_2^2}$
$\quad a(n+1) = a(n) + \rho_a sign\{e(n)\}[(y_1(n) - y_2(n)]\lambda(n)[1 - \lambda(n)]$
$\quad \lambda(n+1) = 1/(1 + e^{-a(n+1)})$

---
$\quad$ % Tracking weight transfer scheme (the proposed method)
$\quad$ if $(mod(n-1, N_0)$ equal to zero)
$\quad\quad$ if $a(n+1) < -a^+$
$\quad\quad\quad a(n+1) = -a^+$
$\quad\quad\quad \lambda(n+1) = 0$
$\quad\quad$ **endif**
$\quad\quad$ if $a(n+1) \geq a^+$
$\quad\quad\quad a(n+1) = a^+$
$\quad\quad\quad \lambda(n+1) = 1$
$\quad\quad\quad \mathbf{w}_2(n+1) = \mathbf{w}_1(n+1)$
$\quad\quad$ **endif**
$\quad$ **endif**
$\quad$ **Let** $n=n+1$
**end**

---

### 2.3 Computational complexity

The computational complexity of the basic CLMS [1], NLMS-NSA [2] and NSA-NSA algorithms is listed in Table 2. Since the basic CLMS combines two LMS algorithms, it requires $4M+2$ multiplications for the adaptation of the component filters. The NLMS-NSA algorithm provides additional insensitivity to the input signal level by combining the NLMS and NSA, it requires 6*M*+1 multiplications for the adaptation of the component filters. In contrast with the CLMS, the proposed algorithm uses NSA as the fast filter to replace the NLMS filter, which reduces the computational burden and the negative effect of impulsive noise. From (1) and (4), the basic-CLMS and NLMS-NSA algorithms need 6 multiplications to compute the filter output and to update $a(n)$. However, the proposed algorithm requires 5 multiplications to update $a(n)$ (see (1) and (6), respectively). According to (9), all the algorithms demand 2*M* multiplications to calculate the explicit weight vector. Moreover, due to using the slide window of tracking weight transfer scheme, the NSA-NSA can further reduce the computation operations. Consequently, these would lead to significant computational efficiency.

**Table 2.** Summary of the computational complexity.

| Algorithms | Component filter adaptation | Basic combination | Explicit weight calculation | Weight transfer |
|---|---|---|---|---|
| Basic-CLMS [1] | 4*M*+2 | 6 | 2*M* | 2*M* |
| NLMS-NSA [2] | 6*M*+1 | 6 | 2*M* | *M*+3 |
| NSA-NSA | 6*M* | 5(using (6)) | 2*M* | No |

## 2.4 The analysis of the mixing parameter

In this section, the convergence behavior of the mixing parameter is analyzed, and the range of $\rho_a$ will be discussed. When the error term $e(n)$ is expanded with a Taylor series [18-20], we have

$$e(n+1) = e(n) + \frac{\partial e(n)}{\partial a(n)} \Delta a(n) + \frac{1}{2}\frac{\partial^2 e(n)}{\partial^2 a(n)} \Delta a^2(n) + h.o.t \quad (10)$$

where *h.o.t.* represents the higher order terms of the remainder of the Taylor series expansion. According to $e(n) = d(n) - y(n)$ and (3), $\frac{\partial e(n)}{\partial a(n)}$ can be obtained as follows:

$$\frac{\partial e(n)}{\partial a(n)} = \lambda(n)[1-\lambda(n)][y_2(n) - y_1(n)]. \quad (11)$$

The mixing parameter correction $\Delta a(n)$ can be calculated from (6)

$$\Delta a(n) = \rho_a sign(e(n))[(y_1(n) - y_2(n))]\lambda(n)[1-\lambda(n)]. \quad (12)$$

Combining (10), (11) and (12), we can express (10) as

$$e(n+1) = e(n)[1 - \frac{\rho_a (y_1(n) - y_2(n))^2}{|e(n)|}\lambda(n)^2(1-\lambda(n))^2]. \quad (13)$$

The NSA-NSA can converge if

$$|e(n+1)| \leq |e(n)| |1 - \frac{\rho_a (y_1(n) - y_2(n))^2}{|e(n)|} \lambda(n)^2 (1-\lambda(n))^2 |. \quad (14)$$

Hence

$$|1 - \frac{\rho_a (y_1(n) - y_2(n))^2}{|e(n)|} \lambda(n)^2 (1-\lambda(n))^2 | < 1. \quad (15)$$

Solving the inequality with respect to $\rho_a$ gives

$$0 < \rho_a < \frac{2|e(n)|}{(y_1(n) - y_2(n))^2 \lambda(n)^2 (1-\lambda(n))]^2}. \quad (16)$$

### 2.5  Steady-state performance of the proposed algorithm

To measure the steady-state performance, the EMSEs of the filters are expressed as [1]

$$J_{ex,i}(\infty) = \lim_{n \to \infty} E\{e_{a,i}^2(n)\}, \quad i = 1, 2 \quad (17)$$

$$J_{ex}(\infty) = \lim_{n \to \infty} E\{e_a^2(n)\} \quad (18)$$

$$J_{ex,12}(\infty) = \lim_{n \to \infty} E\{e_{a,1}(n) e_{a,2}(n)\} \quad (19)$$

where $E\{\cdot\}$ denotes the expectation, $J_{ex,i}(\infty)$ represents the individual EMSE of the $i$th filter, $J_{ex}(\infty)$ is the cross-EMSE of the combined filters, $J_{ex,12}(\infty)$ is the steady-state correlation between the *a priori errors* of the elements of the combination, $e_{a,i}(n)$ and $e_a(n)$ are *a priori error*, respectively, defined by

$$e_{a,i}(n) = [\mathbf{w}_0 - \mathbf{w}_i(n)]^T \mathbf{x}(n) = \varsigma_i^T(n)\mathbf{x}(n) \quad (20)$$

$$e_a(n) = [\mathbf{w}_0(n) - \mathbf{w}(n)]^T \mathbf{x}(n) = \varsigma^T(n)\mathbf{x}(n) \quad (21)$$

where $\varsigma_i(n)$ is the weight error vector of the individual filter, and $\varsigma(n)$ is the weight error vector of the overall filter.

Additionally, for the modified combination (2), $J_{ex,u}(\infty)$ is defined as

$$J_{ex,u}(\infty) = \lim_{n \to \infty} E\{\lambda_u^2(n) e_{a,1}^2(n) + [1-\lambda_u(n)]^2 e_{a,2}^2(n) + 2\lambda_u(n)[1-\lambda_u(n)]e_{a,1}(n)e_{a,2}(n)\} \quad (22)$$

where

$$\lambda_u(n) = \begin{cases} 1 & a(n) \geq a^+ - \varepsilon \\ \lambda(n) & a^+ - \varepsilon > a(n) > -a^+ + \varepsilon \\ 0 & a(n) \leq -a^+ + \varepsilon \end{cases} \quad (23)$$

and $\varepsilon$ is a small positive constant.

Taking expectations of both sides of (6) and using $y_1(n) - y_2(n) = e_{a,2}(n) - e_{a,1}(n)$ yields:

$$E\{a(n+1)\} = E\{a(n)\} + \mu_a E\{sign(e(n))[e_{a,2}(n) - e_{a,1}(n)]\lambda(n)[1-\lambda(n)]\}. \quad (24)$$

According to the Price theorem [21,25], we have

$$E\{sign[e(n)]\theta(n)\} \approx \sqrt{\frac{2}{\pi}} \frac{1}{\chi_{e,n}} E\{e(n)\theta(n)\} \tag{25}$$

where $\chi_{e,n}$ is the standard deviation of the error $e(n)$, i.e., $\chi_{e,n}^2 = E\{e^2(n)\}$, and $\theta(n)$ can be defined as $\theta(n) = [e_{a,2}(n) - e_{a,1}(n)]\lambda(n)[1-\lambda(n)]$. Therefore, (24) becomes

$$E\{a(n+1)\} \approx E\{a(n)\} + \phi_a E\{e(n)[e_{a,2}(n) - e_{a,1}(n)]\lambda(n)[1-\lambda(n)]\}) \tag{26}$$

where $\phi_a = \mu_a \sqrt{\frac{2}{\pi}} \frac{1}{\chi_{e,n}}$. Then, (26) can be rewritten as:

$$E\{a(n+1)\} \approx [E\{a(n)\} + \phi_a E\{[e_{a,2}^2(n) - e_{a,1}(n)e_{a,2}(n)]\lambda(n)[1-\lambda(n)]^2) \\ + \phi_a E\{[e_{a,1}(n)e_{a,2}(n) - e_{a,1}^2(n)]\lambda^2(n)[1-\lambda(n)]\})]_{-a^+}^{a^+}. \tag{27}$$

Assume $\lambda(n)$ is independent of a prior error $e_{a,i}(n)$ in the steady state, under this assumption, $E\{a(n+1)\}$ is governed by

$$E\{a(n+1)\} \approx [E\{a(n)\} + \phi_a E\{\lambda(n)[1-\lambda(n)]^2 \Delta J_2) \\ - \phi_a E\{\lambda^2(n)[1-\lambda(n)]\Delta J_1\}]_{-a^+}^{a^+} \tag{28}$$

where $\Delta J_i = J_{ex,i}(\infty) - J_{ex,12}(\infty), i = 1,2$. Suppose the NSA-NSA convergences, the optimal mean combination weights under convex constraint are given by [1], which is discussed in the three situations as follows:

1) If $J_{ex,1}(\infty) \leq J_{ex,12}(\infty) \leq J_{ex,2}(\infty)$, we have $\Delta J_1 \leq 0$ and $\Delta J_2 \geq 0$. Since $a(n)$ and $\lambda(n)$ are limited in the effective range, an assumption can be expressed as

$$E\{a(n+1)\} \geq [E\{a(n)\} + C]_{-a^+}^{a^+} \text{ as } n \to \infty \tag{29}$$

where $C = \lambda^+(1-\lambda^+)^2(\Delta J_2 - \Delta J_1)$ is a positive constant. In this case, we can conclude that

$$\begin{cases} J_{ex}(\infty) \approx J_{ex,1}(\infty) \\ J_{ex,u}(\infty) = J_{ex,1}(\infty) \end{cases}. \tag{30}$$

Therefore, (30) shows that the NSA-NSA algorithm performs as well as the component filters.

2) If $J_{ex,1}(\infty) \geq J_{ex,12}(\infty) \geq J_{ex,2}(\infty)$, we have $\Delta J_1 \geq 0$ and $\Delta J_2 \leq 0$. Then, (28) can be rewritten as

$$E\{a(n+1)\} \leq [E\{a(n)\} - C]_{-a^+}^{a^+} \text{ as } n \to \infty \tag{31}$$

for a positive constant $C = \lambda^+(1-\lambda^+)^2(\Delta J_1 - \Delta J_2)$ and

$$\begin{cases} J_{ex}(\infty) \approx J_{ex,2}(\infty) \\ J_{ex,u}(\infty) = J_{ex,2}(\infty) \end{cases}. \tag{32}$$

From (32), the overall filter performs approximately equal to the better component filter.

3) If $J_{ex,12}(\infty) < J_{ex,i}(\infty), i = 1,2$, we have $\Delta J_1 > 0$ and $\Delta J_2 > 0$.

Assume $\lambda(n) \to 0$ when $n \to \infty$, we obtain
$$[1-\bar{\lambda}(\infty)]\Delta J_2 = \bar{\lambda}(\infty)\Delta J_1 \tag{33}$$
where $\bar{\lambda}(\infty)$ is given by
$$\bar{\lambda}(\infty) = [\frac{\Delta J_2}{\Delta J_1 + \Delta J_2}]_{1-\lambda^+}^{\lambda^+}. \tag{34}$$

Consequently, it can be concluded from (34) that: if $J_{ex,1}(\infty) < J_{ex,2}(\infty)$, then $\lambda^+ \geq \bar{\lambda}(\infty) > 0.5$; if $J_{ex,1}(\infty) > J_{ex,2}(\infty)$, so $0.5 \geq \bar{\lambda}(\infty) > 1-\lambda^+$.

Consider the following formulas
$$J_{ex}(\infty) = \bar{\lambda}^2(\infty)J_{ex,1}(\infty) + [1-\bar{\lambda}(\infty)]^2 J_{ex,2}(\infty) + 2\bar{\lambda}(\infty)[1-\bar{\lambda}(\infty)]J_{ex,12}(\infty) \tag{35}$$
$$J_{ex,u}(\infty) = \bar{\lambda}^2(\infty)J_{ex,1}(\infty) + [1-\bar{\lambda}(\infty)]^2 J_{ex,2}(\infty) + 2\bar{\lambda}(\infty)[1-\bar{\lambda}(\infty)]J_{ex,12}(\infty) \tag{36}$$
and rearranging (35), we have
$$\begin{aligned} J_{ex}(\infty) &= \bar{\lambda}(\infty)\{J_{ex,1}(\infty) + [1-\bar{\lambda}(\infty)]J_{ex,12}(\infty)\} \\ &\quad + [1-\bar{\lambda}(\infty)]\{[1-\bar{\lambda}(\infty)]J_{ex,2}(\infty) + \bar{\lambda}(\infty)J_{ex,12}(\infty)\} \\ &= \bar{\lambda}(\infty)\{J_{ex,12}(\infty) + \bar{\lambda}(\infty)[J_{ex,1}(\infty) - J_{ex,12}(\infty)]\} \\ &\quad + [1-\bar{\lambda}(\infty)]\{J_{ex,12}(\infty) + [1-\bar{\lambda}(\infty)][J_{ex,2}(\infty) - J_{ex,12}(\infty)]\}. \end{aligned} \tag{37}$$
Then, we can rewrite (37) using (34) as
$$\begin{aligned} J_{ex}(\infty) &= \bar{\lambda}(\infty)[J_{ex,12}(\infty) + \bar{\lambda}(\infty)\Delta J_1] \\ &\quad + [1-\bar{\lambda}(\infty)]\{J_{ex,12}(\infty) + [1-\bar{\lambda}(\infty)]\Delta J_2\} \end{aligned} \tag{38}$$
Since $\bar{\lambda}(\infty) = \Delta J_2/(\Delta J_1 + \Delta J_2)$ and $1-\bar{\lambda}(\infty) = \Delta J_1/(\Delta J_1 + \Delta J_2)$, yielding
$$J_{ex}(\infty) = \bar{\lambda}(\infty)[J_{ex,12}(\infty) + \frac{\Delta J_1 \Delta J_2}{\Delta J_1 + \Delta J_2}] + [1-\bar{\lambda}(\infty)][J_{ex,12}(\infty) + \frac{\Delta J_1 \Delta J_2}{\Delta J_1 + \Delta J_2}]. \tag{39}$$
Hence, we obtain
$$J_{ex}(\infty) = J_{ex,u}(\infty) = J_{ex,12}(\infty) + \frac{\Delta J_1 \Delta J_2}{\Delta J_1 + \Delta J_2}. \tag{40}$$
According to $\bar{\lambda}(\infty) \in (1-\lambda^+, \lambda^+)$, the following bounds hold:
$$J_{ex}(\infty) = J_{ex,u}(\infty) = J_{ex,12}(\infty) + \bar{\lambda}(\infty)\Delta J_1 < J_{ex,1}(\infty) \tag{41}$$
$$J_{ex}(\infty) = J_{ex,u}(\infty) = J_{ex,12}(\infty) + \bar{\lambda}(\infty)\Delta J_2 < J_{ex,2}(\infty). \tag{42}$$
That is
$$\begin{cases} J_{ex}(\infty) < \min\{J_{ex,1}(\infty), J_{ex,2}(\infty)\} \\ J_{ex,u}(\infty) < \min\{J_{ex,1}(\infty), J_{ex,2}(\infty)\} \end{cases}. \tag{43}$$

From the above three situations, it is clear that the proposed NSA-NSA filter performs equally or outperforms the best component filter.

## 3 Simulation results

To evaluate the performance of the proposed algorithm, three examples of system (channel) identification were carried out. The results presented here were obtained from 200 independent Monte Carlo trials. The software of Matlab 8.1 version (2013a) was used to simulate the proposed algorithm under the computer environment of AMD (R) A-10 CPU 2.10 GHz and 8Gb memory. To measure the performance of the algorithms, EMSE using logarithmic scale(dB) was used, defined as:

$$\text{EMSE} = 10\log_{10}\{|e_a^2(n)|\}. \tag{44}$$

The unknown system was a ten-tap FIR filter given by random. White Gaussian noise (WGN) with zero mean and unit variance was used as input. The system was corrupted by additive WGN and an impulsive noise sequence. The impulsive noise $v(n)$ was generated from the Bernoulli-Gaussian (BG) distribution [4,17,23,29]

$$v(n) = A(n)I(n) \tag{45}$$

where $A(n)$ is a binary independent and identically distributed (i.i.d.) Bernoulli process with $p\{A(n)=1\}=c$ and $p\{A(n)=0\}=1-c$, and $c$ is the probability of occurrence for the impulsive interference $I(n)$. The mean value of $v(n)$ is zero, and its variance is given by

$$\text{var}\{v(n)\} = c\sigma_I^2 \tag{46}$$

where $\sigma_I^2 = \text{var}\{I(n)\}$, and the parameters $c$ is set as $c = 0.01$ [4,17,23].

### 3.1 Example 1

For the first example, the parameter $\sigma_I^2$ in (46) was fixed at $\sigma_I^2 = 10^4/12$, and the 10dB SNR WGN [4,17,23]. The unknown system changes abruptly at $n=10000$.

Figs. 4 and 5 show the performances of the proposed algorithm with different sets of $N_0$ and $\rho_a$. The filter values of the NSA were $\mu_1 = 0.05$, $\mu_2 = 0.005$ (which satisfies the stability condition), $\varepsilon_1 = \varepsilon_2 = 0.0001$, and $a^+ = 4$. Consider the stability of evolution of the mixing parameter and the convergence rate, the best choice is $N_0 = 2$. In addition, we can observe from Fig. 5 that the best choice is $\rho_a = 10$.

Figs. 6 and 7 display the evolution of the mixing parameters $\lambda(n)$ and $a(n)$ in NSA-NSA. Run 1 used the no transfer scheme [1], Run 2 and Run 3 represent the mixing parameters based on the tracking weight transfer scheme, according to (4) and (6), respectively. Results demonstrate that the proposed transfer scheme achieves faster convergence rate and improve the filter robustness in the presence of impulsive noise. Moreover, Figs. 6 and 7 shows that adjusting the mixing parameter $a(n)$ using (6) (Run3) results in better stability than other methods.

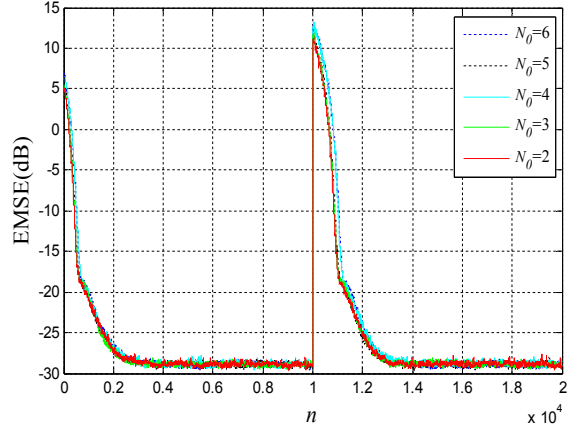

**Fig. 4.** The choice of parameter $N_0$ in example 1.

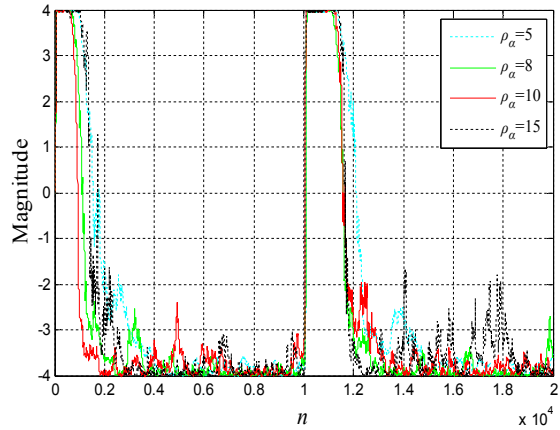

**Fig. 5.** The choice of parameter $\rho_a$ in example 1.(the mixing parameter $a(n)$).

To further show the performance advantage of the proposed method, Fig. 8 depicts the learning curves of the NLMS-NSA and the NSA-NSA algorithms. This figure verifies that the performance of the proposed algorithm is at least as good as the better component in the combination. Both algorithms have the same misadjustment, since the step size of the slow filters are the same. However, the fast filter of the NLMS-NSA is the NLMS, which results in large misadjustment in high background noise environments. Consequently, the NLMS-NSA suffers from higher misadjustment in the initial convergence stage. Fig. 9 plots a comparison of NRMN [23], NSA [10], VSS-NSA [27], VSS-APSA [29], and the proposed algorithm. Clearly, the NSA has a tradeoff between fast convergence rate and low EMSE, while the proposed algorithm shows a good balance between the steady-state error and convergence rate.

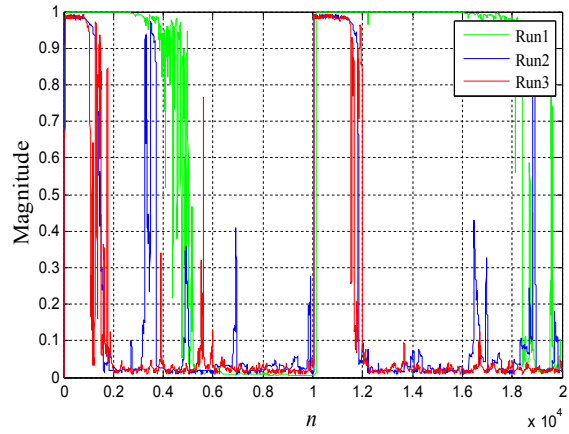

**Fig. 6.** Evolution of the mixing parameter $\lambda(n)$ of NSA-NSA.

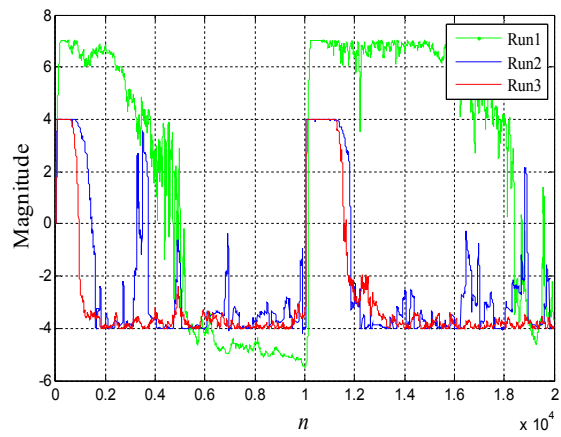

**Fig. 7.** Evolution of the mixing parameter $a(n)$ of NSA-NSA.

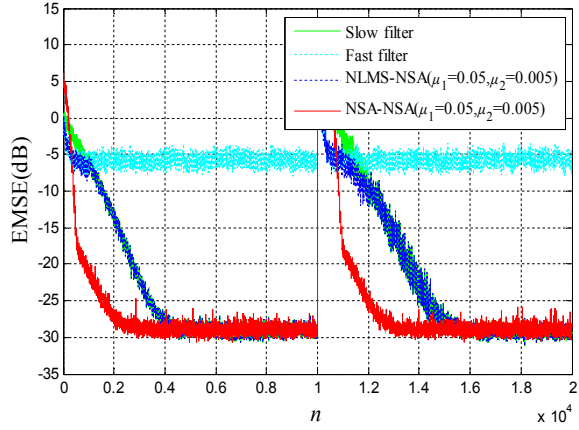

**Fig. 8.** Comparison of EMSE of NLMS-NSA algorithm and NSA-NSA for Gaussian input when 1% impulsive noises are added.

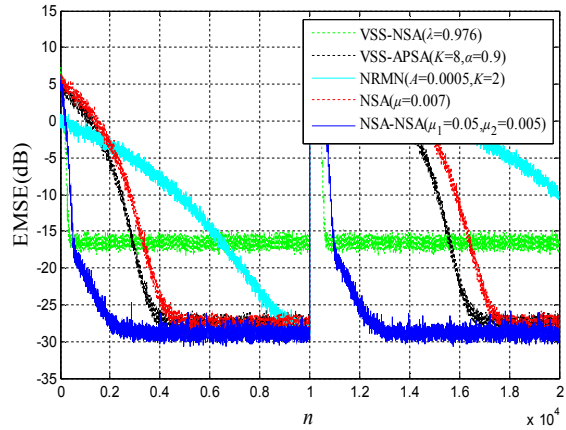

**Fig. 9.** Comparison of EMSE of NRMN, NSA, VSS-NSA, VSS-APSA algorithms and NSA-NSA for Gaussian input when 1% impulsive noises are added.

### 3.2 Example 2

Next, we consider the case of $\sigma_I^2 = 10^4/20$ and SNR=5dB, which corresponds to case with the slightly impulsive case and highly Gaussian noises. The abrupt change appeared in the system at the $n$=10000.

In this example, the step size of NSA-NSA filter was selected as $\mu_1 = 0.05$, $\mu_2 = 0.008$, and $\varepsilon_1 = \varepsilon_2 = 0.0001$. This selection of the parameters ensures good performance of the algorithm in terms of the convergence rate and steady-state misad-justment. Fig. 10 displays the choice of $N_0$ in example 2. We can see that the proposed method is not sensitive to this selection, with the optimal value at $N_0 = 2$. Fig. 11 shows the EMSE of NSA-NSA for different $\rho_a$. The mixing parameter $\rho_a = 10$ for the proposed algorithm was selected to guarantee the stability.

Figs. 12 and 13 show the time evolution of the mixing coefficients, where Run 1 represents the no weight transfer scheme [1], and Run 2 and Run 3 represent the mixing parameters based on the tracking weight transfer scheme given by (4) and (6), respectively. Clearly, it can be observed from these figures that the best selection is Run 3. The robust performance in the presence of impulsive noise is also improved by using (9).

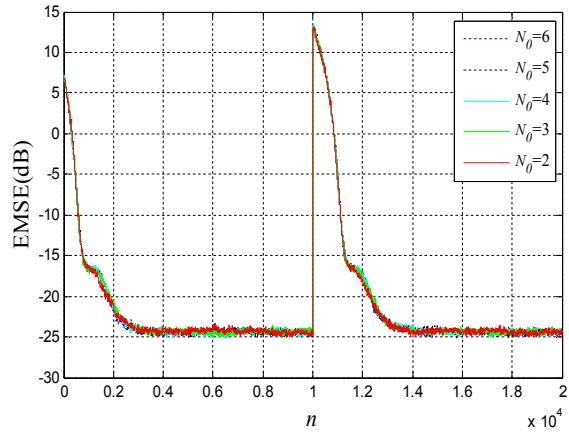

**Fig. 10.** The choice of parameter $N_0$ in example 2.

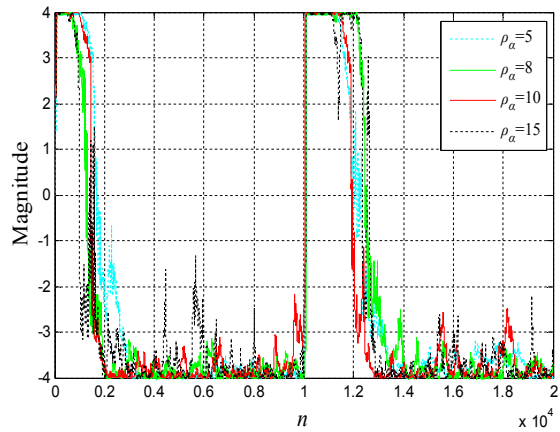

**Fig. 11.** The choice of parameter $\rho_a$ in example 2. (the mixing parameter $a(n)$).

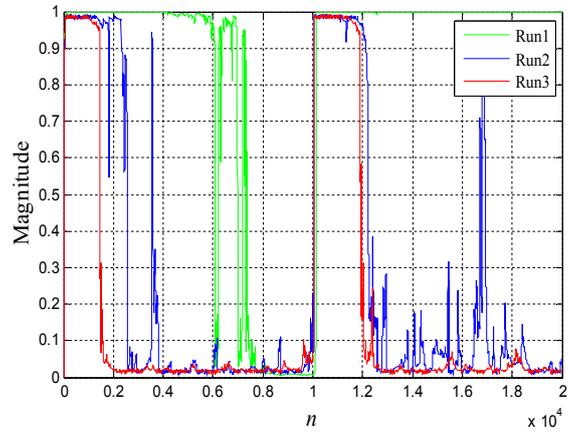

**Fig. 12.** Evolution of the mixing parameter $\lambda(n)$ of NSA-NSA.

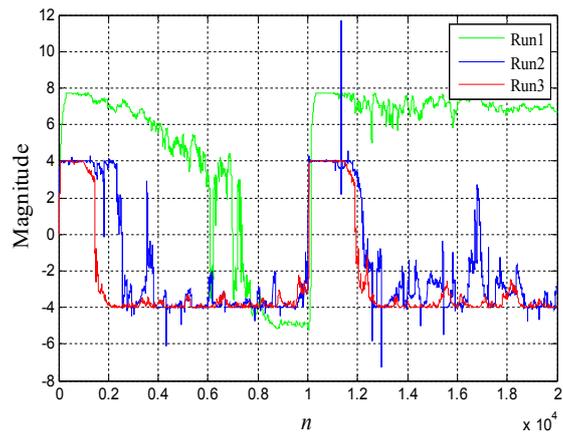

**Fig. 13.** Evolution of the mixing parameter $a(n)$ of NSA-NSA.

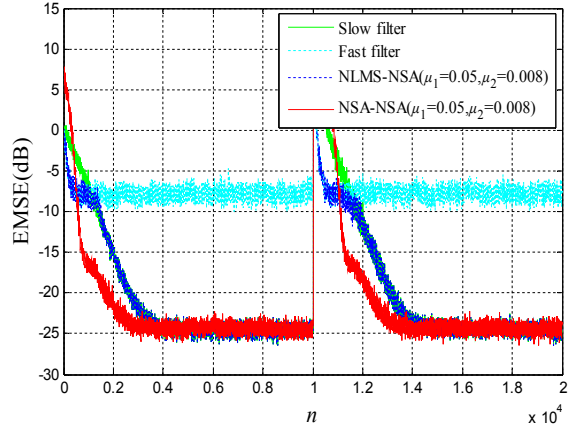

**Fig. 14.** Comparison of EMSE of NLMS-NSA algorithm and NSA-NSA for Gaussian input when 1% impulsive noises are added.

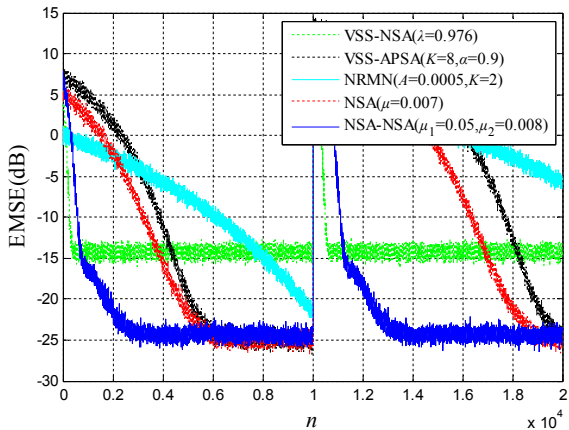

**Fig. 15.** Comparison of EMSE of NRMN, NSA, VSS-NSA, VSS-APSA algorithms and NSA-NSA for Gaussian input when 1% impulsive noises are added.

Fig. 14 plots a comparison of NLMS-NSA and the proposed algorithms. Again, we see that the EMSE of NSA-NSA is consistent with the theoretical analysis. Both algorithms achieve quite similar steady-state error, but the proposed algorithm has the smaller misadjustment in the initial stage of convergence. This is due to the fact that the NLMS algorithm is not well-suited for impulsive noise environment. Fig. 15 shows a comparison of the learning curves from NRMN [23], NSA [10], VSS-NSA [27], VSS-APSA [29] and NSA-NSA for high Gaussian noise and low impulsive noise environments. It is observed that the proposed algorithm achieves an improved performance in the presence of impulsive noise.

## 3.3 Intersymbol Interference (ISI) channel identification under impulsive noise environment

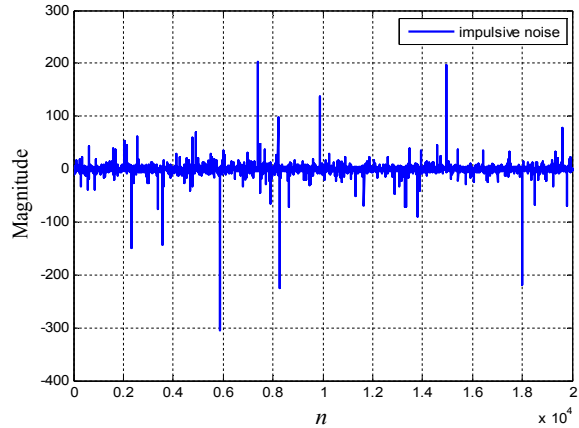

**Fig. 16.** Impulsive noise in ISI channel.

In this section, we consider a real intersymbol interference (ISI) channel corrupted by impulsive noise, which occurs quite often in communication systems. Here, we model the ISI channel as

$$\mathbf{w}_0 = \underbrace{[0.04, -0.05, 0.07, -0.21, -0.5, 0.72, 0.36, 0, 0.21, 0.03, 0.07]^T}_{\text{eleven coefficients}}. \quad (47)$$

In practice, the channel information is unknown. To deal with such problem, the length of our filter was set to $M$=13. Quadrature phase shift keyin (QPSK) was used as the input signal. A segment of 10000 samples was used as the training data and another 10000 as the test data. The ISI channel was corrupted by impulsive noise, as shown in Fig. 16. The performance of the proposed NSA-NSA[1] is demonstrated, in comparison with the NLMS-NSA[2].

Fig. 17 shows the learning curves of the two algorithms in impulsive noise. Clearly, with impulsive noise, the performance of NSA-NSA is barely affected by large disturbances, while the performance of NLMS-NSA deteriorates significantly due to NLMS's sensitivity to outliers.

---

[1] With QPSK input, the adaptation of $a(n)$ of NSA-NSA is given as $a(n+1) = a(n) + \rho_a conj\{sign\{e(n)\}\}[y_1(n) - y_2(n)]\lambda(n)[1 - \lambda(n)]$, where $conj\{\cdot\}$ denotes conjugate operation.

[2] The derivation of VSS-NSA, VSS-APSA, and NRMN are different from the original literatures, when input signal is the complex number. For paper length optimization, and in order to focus on the simplicity of the proposed approach, we have decided to only compare to NLMS-NSA algorithm.

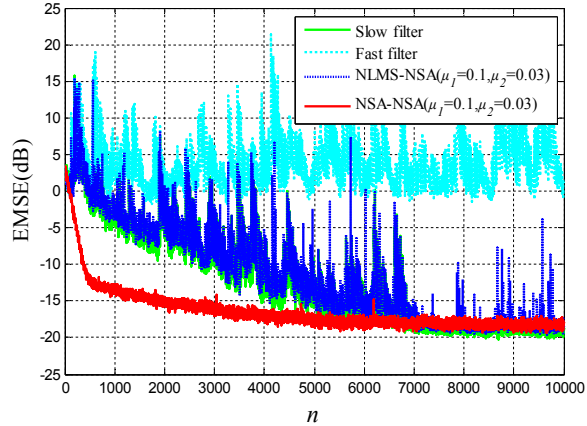

**Fig. 17.** Learning curves of NLMS-NSA and NSA-NSA in ISI channel identification (Testing stage).

## 4    Conclusions

A novel NSA-NSA was proposed to improve the performance of NSA for system identification under impulsive noise. The proposed adaptive convex scheme which combines a fast and a slow NSA filter to achieve both fast convergence speed and low steady-state error. Moreover, a sign cost function scheme to adjust the mixing parameter was introduced to improve the robustness of the algorithm under impulsive noise. To further accelerate the initial convergence rate, a tracking weight transfer scheme was applied in the NSA-NSA. Simulation results demonstrated that the proposed algorithm has better performance than the existing algorithms in terms of convergence rate and steady-state error.

### Acknowledgments

The authors want to express their deep thanks to the anonymous reviewers for many valuable comments which greatly helped to improve the quality of this work.

This work was supported in part by National Natural Science Foundation of China (Grants: 61271340, 61571374, 61134002, 61433011, U1234203), the Sichuan Provincial Youth Science and Technology Fund (Grant: 2012JQ0046), and the Fundamental Research Funds for the Central Universities (Grant: SWJTU12CX026).